# Insight into the Role of Oxygen in Phase-Change Material GeTe

Linggang Zhu[1,2], Zhen Li[1], Jian Zhou[1], Naihua Miao[1,2] and Zhimei Sun[1,2,*]

[1]School of Materials Science and Engineering, Beihang University, Beijing 100191, China
[2]Center for Integrated Computational Materials Engineering, International Research Institute for Multidisciplinary Science, Beihang University, Beijing 100191, China

*Corresponding Author. Email: zmsun@buaa.edu.cn

**Abstract**

Oxygen is widely used to tune the performance of chalcogenide phase-change materials in the usage of phase-Change random access memory (PCRAM) which is considered as the most promising next-generation non-volatile memory. However, the microscopic role of oxygen in the write-erase process, i.e., the reversible phase transition between crystalline and amorphous state of phase-change materials is not clear yet. Using oxygen doped GeTe as an example, this work unravels the role of oxygen at the atomic scale by means of *ab initio* total energy calculations and *ab initio* molecular dynamics simulations. Our main finding is that after the amorphization and the subsequent re-crystallization process simulated by *ab initio* molecular dynamics, oxygen will drag one Ge atom out of its lattice site and both atoms stay in the interstitial region near the Te vacancy that was originally occupied by the oxygen, forming a "dumbbell-like" defect (O-$V_{Te}$-Ge), which is in sharp contrast to the results of *ab initio* total energy calculations at 0 K showing that the oxygen prefers to substitute Te in crystalline GeTe. This specific defect configuration is found to be responsible for the slower crystallization speed and hence the improved data retention of oxygen doped GeTe as reported in recent experimental work. Moreover, we find that the oxygen will increase the effective mass of the carrier and thus increases the resistivity of GeTe. Our results unravel the microscopic mechanism of the oxygen-doping optimization of phase-change material GeTe, and the present reported mechanism can be applied to other oxygen doped ternary chalcogenide phase-change materials.

## 1. Introduction

With the rapid advances of electronic technology and developments of IoT (Internet of Things), non-volatile memories with much better performances than the silicon-based Flash are needed.[1-2] Phase-Change random access memory (PCRAM) shows great promises as the replacement of Flash.[3-7] The working mechanism of PCRAM is based on the contrast properties of crystalline and amorphous state of phase-change material which corresponds to logic state "1" and "0", respectively. GeTe-$Sb_2Te_3$ pseudobinary compounds have been proved to be the most promising phase-change materials due to their fast amorphous-crystalline transition speed, etc.[8-14] Nevertheless, even the best candidate $Ge_2Sb_2Te_5$ among the GeTe-$Sb_2Te_3$ pseudobinary has certain deficiency in PCRAM applications. Therefore, composition optimization by alloying another type of element is an effective way to improve the performance of GeTe-$Sb_2Te_3$ pseudobinary system as the phase-change material.[15-18] In particular, nitrogen and oxygen atoms which are treated as impurity in certain circumstances in the semiconductor industry, exhibit surprisingly beneficial effects on the properties of phase-change materials. Extensive works have been done to explore the effects of oxygen or nitrogen on the properties of phase change material $Ge_2Sb_2Te_5$ (GST) and GeTe.[19-25] Some general conclusions have been made including that oxygen/nitrogen will modify the local structure of amorphous GST/GeTe due to the strong Ge-O/Ge-N bonding. Most importantly, the doping effects are about the crystallization kinetics, for example, it is found that N can slow down the crystal growth of GST,[26-28] which will improve the data-retention property of the amorphous GST. While for GeTe, oxygen was reported to increase its crystallization temperature,[21] thus enabling the application of GeTe at higher temperature. More recently, further experimental study[20] showed that oxygen can also increase the resistivity of GeTe, resulting in a substantial reduction of switching voltage for the devices based on GeTe-O films. Given the substantial influences of oxygen on the structure, crystallization kinetics as well as the electronic properties of GeTe, understanding the microscopic mechanism at atomic scale for these effects is imperative.

GeTe is the first discovered material that demonstrate fast recrystallization,[3, 29] and PCRAM based on GeTe shows fast switching speed and high stability,[30-31] which is thus attractive for the researchers. Moreover, as a prototype phase-change material and the parent compound for the GeTe-$Sb_2Te_3$ pseudobinary phase-change system, GeTe is an ideal system for the research of the fundamental issues in phase-change materials. In the present work, we concentrate on the investigations of the influence of oxygen on properties of GeTe by using atomic scale simulations. Here, both *ab initio* and *ab initio* molecular dynamic calculations (AIMD) simulations are performed to study the crystalline-amorphous-crystalline transition process of oxygen doped GeTe. The fundamental knowledge and insights we aim to obtain include: (1) *Doping position of oxygen.* For the small-size doping element (high mobility), the temperature effect should be determined when studying its doping position in GeTe, given that crystallization temperature of the phase-change material is quite high (above 450 K for GeTe[21]) and the crystallization is finished within a very short time period (within a few nanoseconds). (2) *Microscopic mechanism of the retardation effects of oxygen on the crystallization kinetics of GeTe.* The underling atomic-scale picture will be presented. (3) *Physical mechanism for the conductivity change of GeTe induced by oxygen.* By direct calculations of the conductivity, the structure-conductivity relationship for the oxygen doped GeTe will be unraveled.

2. Methods and calculation details

All the *ab initio* and *ab initio* molecular dynamic calculations are performed by using VASP (Vienna Ab initio Simulation Package).[32-34] The generalized gradient approximation (GGA) parameterized by Perdew and Wang (PW91)[35] is used to describe the electronic exchange-correlation potential.

*Crystalline properties.* For crystalline GeTe, a 54-atom supercell with rhombohedra symmetry is constructed. The cutoff energy of the plane wave basis is set as 450 eV. The doping configurations of oxygen studied here include substituting Ge/Te, occupying interstitial region and staying near a $Ge_{Te}$ antisite defect as the interstitial atom.

*Amorphization process.* The amorphous structure of oxygen doped GeTe is obtained using melt-quench method based on a 200-atom supercell. We adopt the quenching procedure in the literature,[36] where the amorphous structure of pure GeTe is studied: after randomizing the model at 3000K, it is quenched to 1100 K (melting point) within 20ps, then a equilibrium at 1100K for 30ps is followed by a quenching to 300 K over 90 ps.

*Crystallization process.* With the obtained amorphous structure, the crystallization process is investigated by employing the template growth method: the crystalline "template" is bounded to the amorphous structure, resulting in a 300-atom supercell, larger than the previous doped GeTe models.[17, 37] The lattice parameters of the supercell is set based on the experiment measured density of the material; the crystallization simulation is done at 470 K, close to the experiment measured crystallization temperature.[21] Three consecutive simulations, each of which lasts for 180 ps, are performed. The time step of the simulation is 3 fs, and the whole simulations consist of 180000 steps. The AIMD simulations are done by using a cutoff energy of 250 eV and the Brillouin zone is sampled at Gamma point. The snapshots of the atomic structures during the crystallization process are fully relaxed at 0K with the larger cutoff energy of 450 eV, and the relaxed structures are identical to those obtained by using cutoff energy of 250 eV.

*Conductivity calculation.* The conductivity of crystalline GeTe with oxygen was estimated using the semi-classical Boltzmann transport theory within rigid band approximation as implemented the BoltzTraP code.[38]

## 3. Results and Discussions

3.1 Doping position of oxygen in crystalline GeTe: *ab initio* calculations at 0K

Three typical doping positions of oxygen as considered in relevant studies [39] are studied here, i.e., replacing Ge or Te ($O_{Ge}$ or $O_{Te}$), and interstitial site ($O_i$) as shown in Fig. 1. Besides, another defect pair configuration is included, in which oxygen stays in the interstitial site near the Ge that replacing the Te ($Ge_{Te}$). It is worth mentioning that after relaxation the defect pair is switched to what we called "dumbbell-like" defect ($O$-$V_{Te}$-$Ge$): Ge atom deviates from the position of lattice Te after relaxation. The reason why we propose this defect structure is that it is observed in the crystallization simulations of amorphous GeTe which will be discussed in detail in section 3.2.

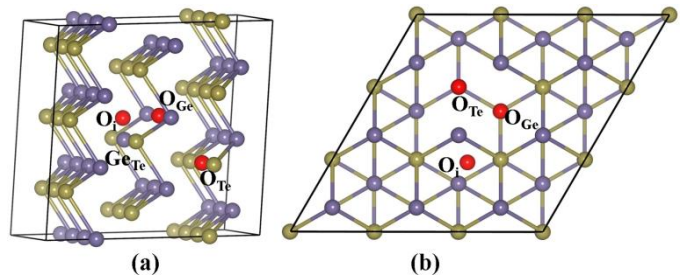

Fig. 1 (a) atomic structure of GeTe supercell with oxygen doping at different sites and (b) its side view. Ge and Te are represented by the purple and yellow balls, while red balls are oxygen atoms.

The most favorable doping position of oxygen in GeTe can be evaluated by the value of the formation energies ($E^f_{defect}$) as shown below.

$$E^f_{O_{Ge}} = E(Ge_{26}Te_{27}O) + \mu(Ge) - E(Ge_{27}Te_{27}) - \mu(O)$$

$$E^f_{O_{Te}} = E(Ge_{27}Te_{26}O) + \mu(Te) - E(Ge_{27}Te_{27}) - \mu(O)$$

$$E^f_{O_i} = E(Ge_{27}Te_{27}O) - E(Ge_{27}Te_{27}) - \mu(O)$$

$$E^f_{O_i + Ge_{Te}} = E(Ge_{28}Te_{26}O) + \mu(Te) - E(Ge_{27}Te_{27}) - \mu(O) - \mu(Ge)$$

In the above equations, E represents the energy of the supercell whose composition is shown in the parenthesis. μ is the chemical potential of the relevant atoms. It is not easy to obtain the absolute value of the above formation energies, since the chemical potential of Ge, Te and O are all variables. Given that the relative values of these energies are needed in our study, here the formation energy of interstitial oxygen $E^f_{O_i}$ is used as the reference and the formation energy difference among various defects can be calculated using the following equations.

$$E^f_{O_{Te}} - E^f_{O_i} = E(Ge_{27}Te_{26}O) + \mu(Te) - E(Ge_{27}Te_{27}O)$$

$$E^f_{O_{Ge}} - E^f_{O_i} = E(Ge_{26}Te_{27}O) + \mu(Ge) - E(Ge_{27}Te_{27}O)$$

$$E^f_{O_i + Ge_{Te}} - E^f_{O_i} = E(Ge_{28}Te_{26}O) + \mu(Te) - E(Ge_{27}Te_{27}O) - \mu(Ge)$$

And now there are two variables left, μ(Ge) and μ(Te). In GeTe, chemical potential of Ge and Te is correlated and restricted in a specific range.

$$E^f_{GeTe} + E(Te_{bulk}) \leq \mu(Te) \leq E(Te_{bulk})$$

$$E^f_{GeTe} = \frac{1}{27}(E(Ge_{27}Te_{27}) - 27E(Ge_{bulk}) - 27E(Te_{bulk}))$$

$$\mu(Ge) = \frac{1}{27}E(Ge_{27}Te_{27}) - \mu(Te)$$

Where $E^f_{GeTe}$ is the formation energy of GeTe compound, $E(Ge_{bulk})$ and $E(Te_{bulk})$ are the energy of each Ge and Te atom in the bulk materials.

Therefore, the formation energy difference among various defect structures can be obtained as the function of chemical potential of Te, as shown in Fig. 2. It can be seen that the oxygen prefer to replace Te compared to other doping positions, similar as other studies in the Ge-Sb-Te system.[39] The main reason should be that both oxygen and Te are VIA elements, thus close chemical properties are expected. It is also interesting to see that the formation of the "dumbbell-like" defect becomes more favorable than the isolated interstitial O in the Te-rich region.

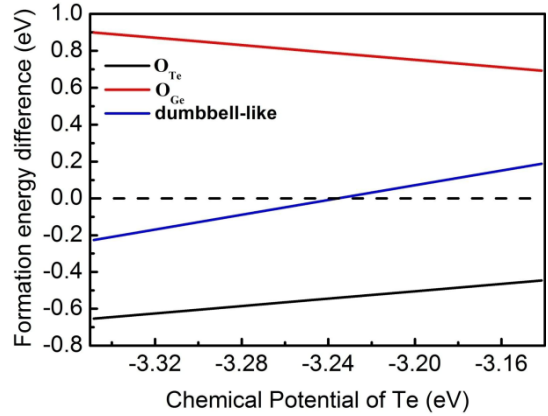

Fig. 2 Formation energies of various defects relative to that of the interstitial oxygen (horizontal dash-line). The negative value means that under specific chemical potential of Te this defect structure is more stable than that with oxygen in the interstitial region.

3.2 Evolution of oxygen configuration in GeTe in the phase-change process

Given the most favorable doping position of O, i.e., occupying position of Te according to the *ab initio* calculations, a crystalline supercell consists of 100 Ge, 99 Te and 1 oxygen is constructed. Then with the melt-quench methods as described in Section 2, the amorphous structure is obtained, which is shown in Fig. 3(a). And afterwards, crystalline GeTe is "bounded" to the obtained amorphous structure (shown in Fig. 3(b)), acting as "seeds" for crystallization process. Our whole crystallization simulation lasts for 540 ps. The movie of the simulation over the first 180 ps can be found in the supplemental material (Movie1). From the energy curve of the system and mean square displacement (MSD) in the initio/final period of the simulations as shown in Fig. 4, it can be seen that the system has reached the equilibrium. Considering such a long time that the system has been equilibrated, we think that the crystallization process can be seen as "completed".

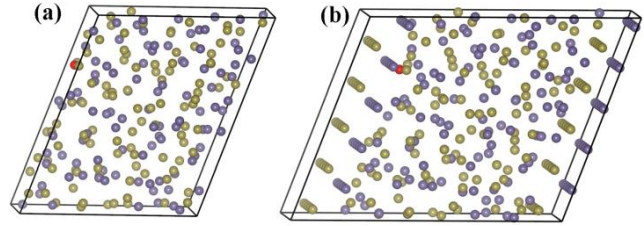

Fig. 3 (a) Atomic structure of amorphous GeTe with oxygen, (b) the amorphous matrix that has been bounded to the crystalline template. Ge and Te are represented by the purple and yellow balls, while red balls are oxygen atoms.

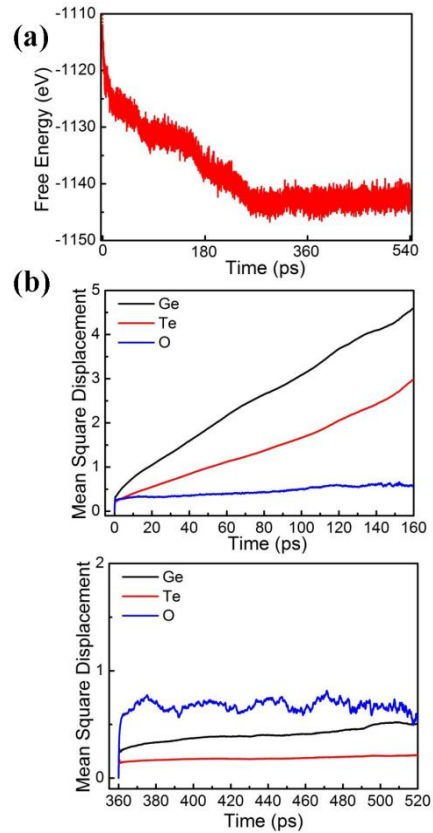

Fig. 4 (a) Energy variance of the GeTe-O system in the crystallization process, (b) mean square displacement (MSD) in the initial (upper figure) and final (lower figure) stage of the crystallization

Based on the simulations of the whole phase-change process, i.e., crystalline-amorphous-crystalline transition, we noticed that the doping position of oxygen and its coordination environment is evolving, as summarized in Fig. 5. Firstly, in the initial input crystalline structure, oxygen occupies the lattice position of Te, which is most energy favorable configuration as shown in Fig. 2. As a substitution of Te, oxygen is surrounded by six Ge (Fig. 5(a)). Among these Ge atoms, three Ge atoms are closer to oxygen than the others three, and the distance between O and Ge are 2.86 and 3.24 Å, respectively. After relaxation, we found that oxygen moves further towards the three nearer neighbors with the distance of 2.05 Å, staying above the triangle formed by three Ge atoms. In the amorphous GeTe, oxygen located in the center of the triangle formed by three Ge atoms (Fig. 5(b)), and the distance between O and Ge are within 2.10 Å. This local triangular structure is also found in the carbon doped amorphous GeTe.[37] In the re-crystallized GeTe (Fig. 5(c), structure at 540 ps of the crystallization simulations), the "$Ge_3O$" triangular configuration is retained, while meanwhile, one of the Ge atom is close to the Te-atom chain and forms a "dumbbell-like" (O-$V_{Te}$-Ge) defect with oxygen. In this $Ge_3O$ structure, the distance between oxygen and Ge are 1.88 to 2.08 Å. Given that the geometry of the Ge-O cluster in the amorphous states are retained in the re-crystallized GeTe, i.e., O can change the local structure of both the amorphous and crystalline state, it shows that the bonding between Ge and O is very strong.

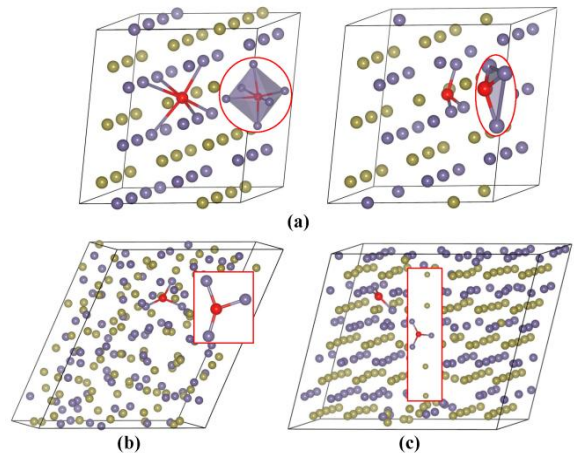

Fig. 5 Oxygen in different states of GeTe. The local coordination environments are heighted in the insets. (a) when oxygen replaces Te in crystalline GeTe, structures before (left) and after relaxation (right) are shown, (b) oxygen is in the center of the triangle formed by three Ge atoms in amorphous GeTe, (c) in re-crystallized GeTe

after the 540 ps annealing simulations, a "dumbbell-like" defect is found. Ge and Te are represented by the purple and yellow balls, while red balls are oxygen atoms.

It is surprising to see that in the re-crystallized GeTe (Fig. 5(c)), the configurations that oxygen replaces Te in the input crystalline GeTe (Fig. 5 (a)) is not recovered. Instead, oxygen comes to the interstitial region, and meanwhile one Ge atom is "dragged" to the position close to the Te vacancy where oxygen was. And in fact, this "dumbbell-like" defect is formed at a quite early stage of the crystallization (at time of 50 ps of the simulations), and remains in the long following annealing process. This abnormal defect configuration is also found in the other independent annealing simulation in our study, indicating the stability of this defect. As shown in Fig. 2, this defect configuration has a higher formation energy according to the calculations at 0 K, thus its occurrence in the crystallization process should be attributed to the temperature effect. To further confirm the stability of the "dumbbell-like" defect configuration at high temperature, a crystalline GeTe supercell with one oxygen substituting Te is annealed at 1100K for 9 ps, and the movie of the simulation can be found in the supplemental materials (Movie2). It is noticed that at such a high temperature, oxygen will not stay in the Te atom chain, and instead it drifts quite freely in local region. More importantly, we found that one Ge atom that is near the oxygen will diffuse into the Te atom chain. A snapshot at 9 ps is shown in Fig. 6, it can be seen that oxygen goes to the interstitial site and meanwhile drags one Ge atom into the Te lattice. After the relaxation at 0K, we found that the "dumbbell-like" defect mentioned above appears. And the ground state energy of this configuration is 0.28 eV higher than the structure with oxygen substitution of Te (leftmost picture in Fig. 6) at 0 K. Therefore, considering the application environment (moderate temperature) of the phase-change material GeTe, it can be expected that this quite stable "dumbbell-like" defect configuration can not be excluded in the crystalline state GeTe (set state).

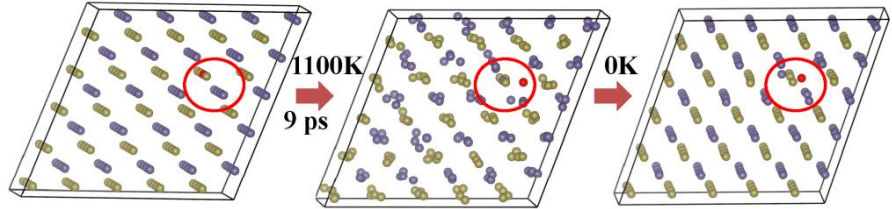

Fig. 6 Structure evolution of oxygen doped GeTe at 1100 K for 9 ps, followed by a relaxation at 0 K. The red circle highlights the significant structure change near the oxygen atom. Ge and Te are represented by the purple and yellow balls, while red balls are oxygen atoms.

3.3 Effects of oxygen on the crystallization kinetics of GeTe

For the template-growth method used in the crystallization simulation process, the atomic structure of the "interface" between the amorphous matrix and crystalline template will affect the crystallization kinetics, resulting in unexpected influence when the effects of oxygen are considered. To exclude these artificial effects, here we use identical input structure to perform the crystallization simulations for GeTe and GeTe-O: oxygen in the GeTe-O (input structure used in the crystallization simulation in section 3.2) is replaced by Te, as shown in Fig. 7 at 0 ps. Therefore the artificial effects from the "interface" between the amorphous and crystalline state are minimized, and any different crystallization-behavior between the GeTe-O and GeTe structure in Fig.7 should be induced by the doping of oxygen.

Snapshots from the simulation process are shown in Fig. 7. Clearly, with time elapse, the arrangements of the atoms in both GeTe-O and GeTe become ordered. For GeTe-O system (Fig. 7(a)), it can be seen that the system have reached the equilibrium after 360 ps, since for the time period from 360 ps to 540 ps, the structure is not changed much, in consistence with the energy variance and MSD curve in Fig. 4. As mentioned above, the oxygen goes to the interstitial region and the "dumbbell-like" defect remains in the whole simulation process. On the other hand, for GeTe (Fig. 7(b)), the whole structure at 180 ps almost becomes crystalline, much more ordered than GeTe-O at 180 ps. This contrast crystallization behavior between GeTe-O and GeTe which can be traced back to the oxygen doping, is clearly seen.

This finding agrees very well with the experimental results that oxygen doping slow down the crystallization speed and increases the crystallization temperature of GeTe[21]. Moreover, as can be seen from the GeTe-O structures at 360 ps and 540 ps, the disordered region is connected to the "dumbbell-like" defects. Therefore, according to our simulations, the microscopic mechanism of the effects of oxygen on the crystallization of GeTe comes from the defects induced by oxygen. The corresponding radio distribution functions for these snapshots are shown in Fig. S1 in the supplemental materials, which show ordering arrangements of the atoms at different stage of the simulations and further confirm the contrast crystallization behavior between GeTe-O and GeTe system.

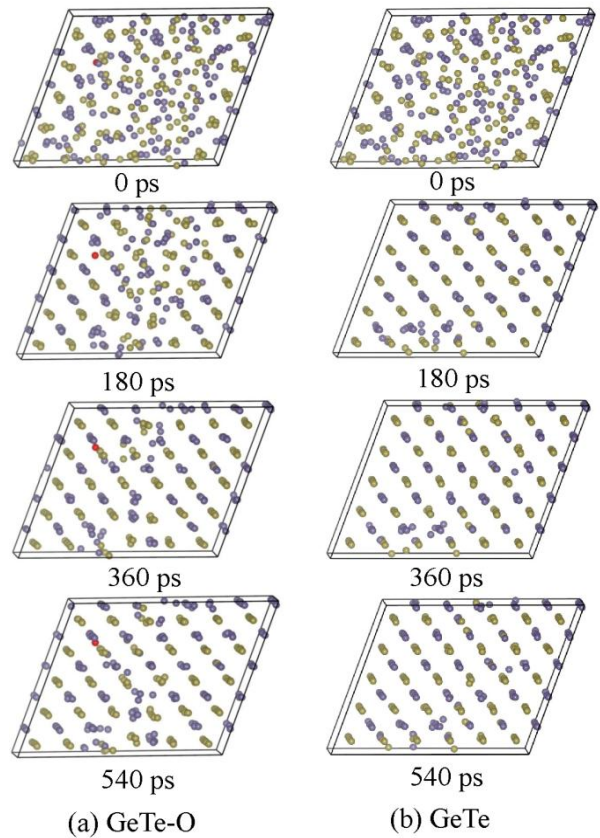

Fig. 7 Snapshots from the crystallization process of (a) GeTe-O system and (b) GeTe system. Ge and Te are represented by the purple and yellow balls, while red ball in (a) is oxygen atom. All configurations have been relaxed at 0K to remove thermal-fluctuation effects for reasonable comparison.

One thing should be pointed out is that for pristine GeTe system, even the system has been reached equilibrium for a long time (more than 360 ps), a "perfect" crystalline structure is not obtained and some point defects still exist after the simulations. In fact, for our template-growth methods, two "grains" are growing from the two boundaries of the supercell into the center, and the point-defect regions are actually near the "grain boundaries" of the two growing grains, making it more difficult to form a perfect crystalline. It is very likely that the crystallization can hardly be "fully" completed within several hundreds of picoseconds. Higher temperature and/or much longer time that beyond the time scale of typical AIMD simulation might be required in order to get a "perfect" crystalline. Nevertheless, the effects of oxygen on the crystallization of GeTe can still be concluded by comparing the crystallization process of GeTe-O and GeTe as clearly shown in Fig. 7.

3.4 Conductivity variance induced by oxygen

In addition to the structural influence of oxygen on GeTe, its effects on the conductivity of GeTe which is closely related to the practical application is also vital and worth studying. The conductivity $\sigma$ is defined as follows.

$$\sigma = ne^2\tau/m^*$$

Where $n$ is the concentration of free carriers with charge $e$, $\tau$ is the average relaxation time, and $m^*$ is the effective mass of the carrier. Given the band structure from the *ab initio* calculations, effective mass $m^*$ which is related to the derivative of the energy bands can be calculated. Then, the relationship between $\sigma/\tau$ and $n$, can be obtained.

Dependence of $\sigma/\tau$ on carrier concentration $n$ at 300 K is shown in Fig. 8. To verify our calculations, the conductivity of pure GeTe at room temperature is estimated: from the experimental data in the literature, the relaxation time $\tau$ at 300 K is 5.3 fs,[40] and the carrier concentration is $21.4 \times 10^{20}$ cm$^{-3}$,[41] then our calculated conductivity is $3.98 \times 10^5$ Sm$^{-1}$. On the other hand, the experimental measured conductivity of pure GeTe by Bahl et al.[40] and Nath et al.[42] is $6.67 \times 10^5$ Sm$^{-1}$ and $2.09 \times 10^5$ Sm$^{-1}$, respectively. Therefore, our theoretical result is in reasonable agreement with the experimental data in the case of pure GeTe.

As shown in Fig. 8, oxygen will decrease the value of $\sigma/\tau$ of GeTe for a certain carrier concentration $n$, independence of its doping position. This means that effective mass is larger in oxygen doped GeTe than in pure GeTe. In Fig. 9, the band structures of GeTe-O systems are shown, and it can be seen that the bands near the Fermi level are narrower and flatter compared to pure GeTe, indicating the increasing of effective mass. Zhou et al[20] suggests that introduction of Ge-O bonds in GeTe will decrease the carrier concentration $n$, leading to the decrease of conductivity. As for the relaxation time $\tau$ representing the interval between scattering events coming from vibrations or collisions with defects, it can be expected that in the oxygen doped GeTe, $\tau$ is smaller than in pure GeTe. Above all, as a results of the decreasing $n$ and $\tau$, combing the increasing of $m^*$ caused by the introduction of oxygen as found here, the conductivity of GeTe-O will be decreased. So far, the origin of the increasing resistivity in GeTe-O compared to pure GeTe as found in the experiments is uncovered.

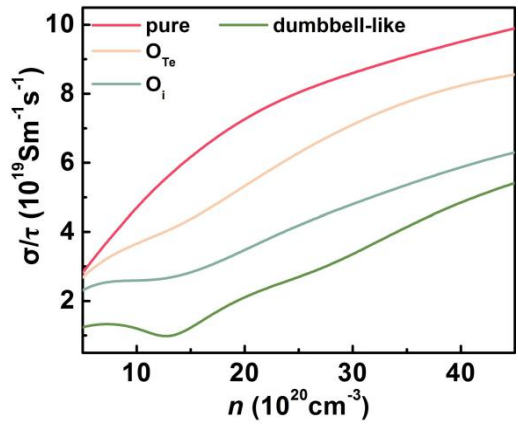

Fig. 8 Calculated relationships between $\sigma/\tau$ and carrier concentration $n$ for GeTe with oxygen doped at different sites at 300K.

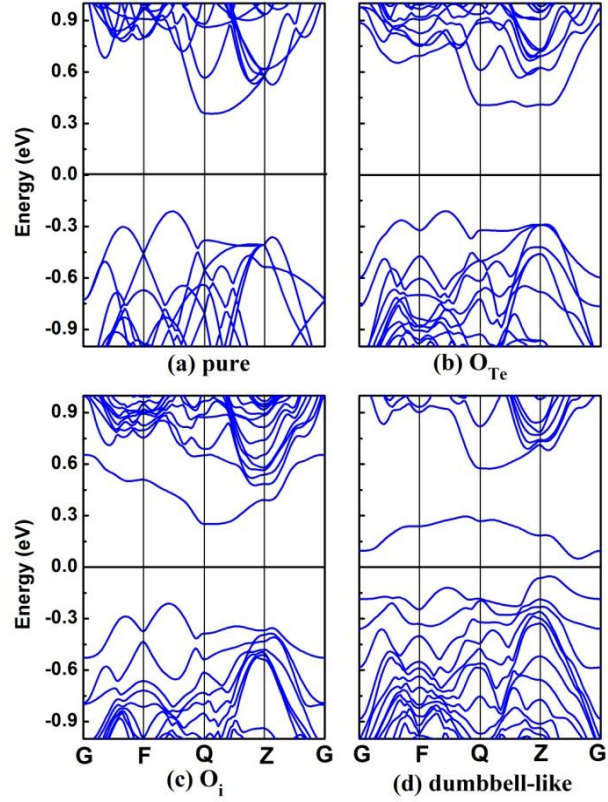

Fig. 9 Band structures for GeTe with oxygen at different doping sites.

## 4　Conclusions

In the present work, crystalline-amorphous-crystalline transition of oxygen doped GeTe, corresponding to the "reset" and "set" process of the PCRAM, are studied by large scale *ab initio* molecular dynamics simulations. Two interesting phenomena are observed in the simulations of the crystallization process. One is that doping position of oxygen (replacing Te) in the input crystalline GeTe can not be recovered after amorphization and re-crystallization process, while instead, a "dumbbell-like" defect (O-$V_{Te}$-Ge), consisting of one Te vacancy accompanied by one interstitial oxygen and one interstitial Ge, is formed. This "dumbbell-like" defect also occurs when the crystalline GeTe with oxygen occupying Te lattice is annealed at high temperature. Therefore, for the doping position of oxygen, the discrepancy between the *ab initio* total energy calculations and *ab initio* molecular dynamic simulations stems from the temperature effects. Secondly, by comparing the crystallization process of GeTe-O and GeTe, it is concluded that oxygen can retard the crystallization of GeTe, and some disordered regions which are connected to the "dumbbell-like" defect remain in the GeTe-O after the simulations. This explains the experimental finding that oxygen increases the crystallization temperature of GeTe. Finally, we find that oxygen can increase the effective mass of the carriers in GeTe, which contributes to the increased resistivity of the crystalline GeTe as observed in the experiments.

**Acknowledgements**

This work is partially supported by National Natural Science Foundation for Distinguished Young Scientists of China (51225205) and the National Natural Science Foundation of China (61274005, 51401009).

**Figure captions:**

Fig. 1 (a) atomic structure of GeTe supercell with oxygen doping at different sites and (b) its side view. Ge and Te are represented by the purple and yellow balls, while red balls are oxygen atoms.

Fig. 2 Formation energies of various defects relative to that of the interstitial oxygen (horizontal dash-line). The negative value means that under specific chemical potential of Te this defect structure is more stable than that with oxygen in the interstitial region.

Fig. 3 (a) Atomic structure of amorphous GeTe with oxygen, (b) the amorphous matrix that has been bounded to the crystalline template. Ge and Te are represented by the purple and yellow balls, while red balls are oxygen atoms.

Fig. 4 (a) Energy variance of the GeTe-O system in the crystallization process, (b) mean square displacement (MSD) in the initial (upper figure) and final (lower figure) stage of the crystallization

Fig. 5 Oxygen in different states of GeTe. The local coordination environments are heighted in the insets. (a) when oxygen replaces Te in crystalline GeTe, structures before (left) and after relaxation (right) are shown, (b) oxygen is in the center of the triangle formed by three Ge atoms in amorphous GeTe, (c) in re-crystallized GeTe after the 540 ps annealing simulations, a "dumbbell-like" defect is found. Ge and Te are represented by the purple and yellow balls, while red balls are oxygen atoms.

Fig. 6 Structure evolution of oxygen doped GeTe at 1100 K for 9 ps, followed by a relaxation at 0 K. The red circle highlights the significant structure change near the oxygen atom. Ge and Te are represented by the purple and yellow balls, while red balls are oxygen atoms.

Fig. 7 Snapshots from the crystallization process of (a) GeTe-O system and (b) GeTe system. Ge and Te are represented by the purple and yellow balls, while red ball in (a) is oxygen atom. All configurations have been relaxed at 0K to remove thermal-fluctuation effects for reasonable comparison.

Fig. 8 Calculated relationships between $\sigma/\tau$ and carrier concentration $n$ for GeTe with oxygen doped at different sites at 300K.

Fig. 9 Band structures for GeTe with oxygen at different doping sites.